\documentclass[aip,apl,twocolumn,superscriptaddress,numerical,preprint]{revtex4}
\usepackage{amssymb,amsmath}
\usepackage{graphicx}
\usepackage{dcolumn}
\usepackage{multirow}
\usepackage{color}
\usepackage{hyperref}
\usepackage{epsfig}
\usepackage{bm}

\begin{document}

\title{\textbf{\fontfamily{phv}\selectfont Ambipolar quantum dots in intrinsic silicon}}
\author{A.~C.~Betz}
\email{ab2106@cam.ac.uk}
\affiliation{Hitachi Cambridge Laboratory, J. J. Thomson Avenue, Cambridge CB3 0HE, United Kingdom}
\author{M.~F.~Gonzalez-Zalba}
\affiliation{Hitachi Cambridge Laboratory, J. J. Thomson Avenue, Cambridge CB3 0HE, United Kingdom}
\author{G. Podd}
\altaffiliation{current address: Base4 Innovation Ltd, Broers Building, J. J. Thomson Avenue, Cambridge CB3 0FA, United Kingdom}
\affiliation{Hitachi Cambridge Laboratory, J. J. Thomson Avenue, Cambridge CB3 0HE, United Kingdom}
\author{A.~J.~Ferguson}
\affiliation{Cavendish Laboratory, University of Cambridge, Cambridge CB3 0HE, United Kingdom}

\date{\today}

\begin{abstract}
We electrically measure intrinsic silicon quantum dots with electrostatically defined tunnel barriers. The presence of both $p$-type and $n$-type ohmic contacts enables the accumulation of either electrons or holes. Thus we are able to study both transport regimes within the same device. We investigate the effect of the tunnel barriers and the electrostatically defined quantum dots. There is greater localisation of charge states under the tunnel barriers in the case of hole conduction leading to higher charge noise in the $p$-regime. 
\end{abstract}

\maketitle

The bandstructure of silicon is asymmetric with respect to the conduction and valence bands and, as a result, the electrical transport characteristics of electrons and holes differ. This is readily observed in the room temperature behaviour of bulk field effect transistors (FETs), where the electron mobility is typically several times higher than the hole mobility \cite{Kittel}. The band asymmetry also affects nanoscale quantum electrical devices in which case the microscopic properties of the carriers, such as effective mass and spin, are the relevant characteristics. These microscopic properties are key factors in the development of future nano-electronic devices, which further highlights the importance of understanding the effect of band asymmetry on the nanoscale \cite{Kane1998,Hanson2008}. 
\newline
In this Letter we investigate the asymmetry of electrons and holes in an intrinsic silicon nanostructure. An ambipolar device design allows us to induce either electrons or holes in an intrinsic silicon metal-oxide-semiconductor (MOS) device, as well as electrostatically defined quantum dots therein, and to study both transport regimes within the same device. We then study the response of both the electrostatically tunable tunnel barriers and the quantum dot to the ambipolar current.

Electrostatically defined $n$-type quantum dots have been measured in doped silicon-on-insulator based fin-FETs using a number of gate materials \cite{Fujiwara2006,Fujiwara2004,GonzalezZalba2012,Roche2012,Prati2012,Voisin2014,Betz2014,GonzalezZalba2014} as well as in intrinsic silicon \cite{Angus2007,Podd2010,Fuechsle2010} and silicon-germanium heterostructures \cite{Simmons2007}. Spin blockade has been demonstrated in these architectures \cite{Liu2008,Rokhinson2001,Shaji2008,Lai2011,Borselli2011} and electron spin relaxation times of the order of seconds \cite{Morello2010,Xiao2010,Simmons2011} have been reported. Given the long spin relaxation time and the reported coherence \cite{Maune2012}, silicon quantum dots are viewed to be one of the most promising platforms for quantum information processing \cite{Morton2011,Zwanenburg2013}. Furthermore, Coulomb blockade through multiple quantum states has been demonstrated at room temperature \cite{Lee_NanoLett2014}.

The $p$-type regime has been studied in electrostatically defined quantum dots \cite{Li2013,Spruijtenburg2013}, silicon-on-insulator fin-FETs \cite{Fujiwara2001,Ono2007} and also in the few hole limit in silicon-based heterostructures and nanowires \cite{Klein2007,Zong2005,Zwanenburg2009}. \newline
Furthermore, Coulomb blockade of both electrons and holes has previously been observed in the disordered channels of nanoscale silicon field effect transistors with both $p$ and $n$-type ohmic contacts \cite{Ishikuro1999}.

\begin{figure}[htbp]
		\includegraphics[width=0.9\columnwidth]{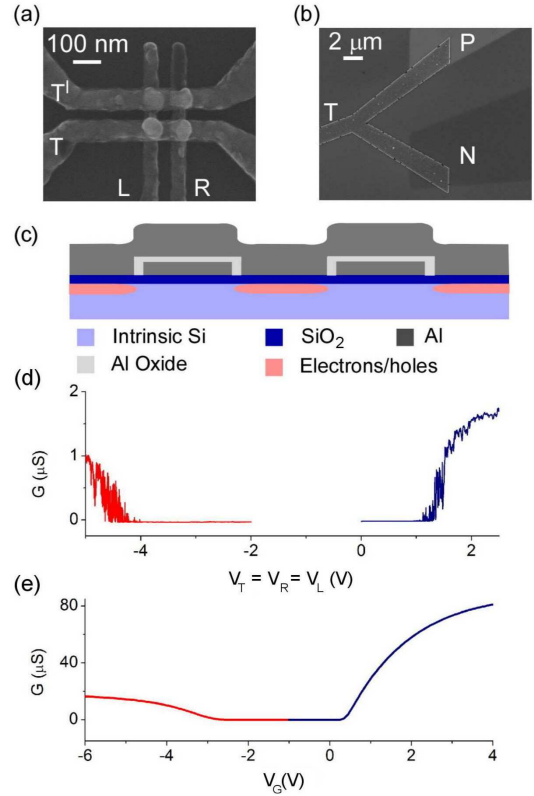}
	\caption{(Color online) (a) Scanning electron micrograph of the sample showing both aluminium gate layers, which in this geometry define two quantum dots between gates L and R. Both dots were operational but were only studied independently. (b) SEM image of the ohmic contact region. The top gate (T) overlaps heavily doped $p$ and $n$ regions allowing either electrons or holes to be introduced into the sample. (c) Schematic cross-section of the device. The topmost gate (T and T') accumulate carriers, while the next lower gate level provides the tunnel barriers. (d) Conductance measurement in nanoscale device at $T\simeq 100\;\mathrm{mK}$ and $V_{ds} = 250\;\mathrm{\mu V}$ of both the $p$ and $n$ regions. All gates are held at the same potential. (e) Conductance measurement of a larger FET with only one topgate in which the channel measures $W\;\times\;L=10\;\times\;80\;\mathrm{\mu m^2}$ at $V_{ds}=10\;\mathrm{mV}$. }
	\label{Fig1}
\end{figure}

Our sample processing starts with a high resistivity ($\rho > 5000\;\mathrm{\Omega\, cm}$) silicon wafer. Boron and phosphorus ohmic contacts are formed by ion implantation ($10^{15}\;\mathrm{cm^{-2}}$ at 7 keV for Boron and 15 keV for Phosphorus) and the implantation damage is repaired by rapid thermal annealing ($1050^{\circ}\;\mathrm{C}$, 15 s). A 15 nm thermal oxide is subsequently grown on the wafer surface by dry oxidisation. After the silicon processing, an electrostatic gate structure is patterned by electron beam lithography. Following the successful "Angus architecture" \cite{Angus2007}, the gates are made from thermally evaporated aluminium and consist of two layers which are electrically isolated by the growth of a native aluminium oxide (Fig. \ref{Fig1}(a) and (c)). The aluminium oxide is formed by heating the lower layer of gates on a hotplate in air ($150^{\circ}\;\mathrm{C}$, 10 minutes), the upper gate layer is subsequently patterned. Typically this oxide provides at least 4 V of electrical isolation between the gate layers. Crucially the upper gate layer overlaps both the n and p-type ohmic contacts allowing both electron and holes to be induced (Fig. \ref{Fig1}(b)). Bond pads made from Al:Si (Si - 1 $ \% $) alloy are thermally evaporated onto the ohmic contacts followed by a post-metallisation anneal at $350^{\circ}\;\mathrm{C}$ in $N_2$ gas.

Throughout this Letter, we present direct current electrical transport measurements with the sample cooled to $T\simeq100\;\mathrm{mK}$ in a dilution refrigerator. The device consists of two upper gates (T and T') and a set of lower gates (L and R), as shown in the micrograph Fig.\ref{Fig1}(a). Gates T and T' induce the electron or hole gas, as shown in Fig.\ref{Fig1}(c), while gates L and R provide the tunnel barriers between quantum dot and leads. Each of the upper gates corresponds to a separate quantum dot sample, which were both investigated during the course of the experiment. As they showed very similar results, we restrict the presentation to data of the bottom device.

We start by characterising a large micro-wire FET of size $W\;\times\;L=10\;\mathrm{\mu m}\;\times\;80\;\mathrm{\mu m}$ measured at liquid helium temperature ($4.2\;\mathrm{K}$) and source-drain voltage $V_{ds}=10\;\mathrm{mV}$. The device is operated as FET, i.e. all gates are held at the same voltage, forming one large gate electrode. The threshold voltages of electron and hole conduction differ with $V_T\simeq0.36\;\mathrm{V}$ for $n$-type and $V_T\simeq-2.8\;\mathrm{V}$ for $p$-type, respectively, representing the aforementioned electron-hole asymmetry. From the linear regions of the $G_{ds}(V_g)$ data in Fig.\ref{Fig1}d we extract mobilities of $\mu_e\simeq8900\;\mathrm{cm^2/Vs}$ and $\mu_h \simeq 1600\;\mathrm{ cm^2/Vs}$ for the $n$ and $p$ regime, respectively, which is broadly in agreement with previously reported low temperature mobilities in silicon \cite{Kawaguchi1981}. The asymmetry may in part be explained by the fact that the interaction time for scattering at impurities or trapped oxide charges is longer for holes due to their larger mass and lower saturation velocity.\newline
We now turn to the scaled down device ($W\;\times\;L\simeq 100\;\mathrm{nm}\;\times 100\;\mathrm{nm}$) and investigate the nano-scale transistor turn-on in both the $p$ and $n$ conduction regime. The upper and barrier gates are held at the same potential, thus forming a nano-wire FET. Fig. \ref{Fig1}(d) shows the conductance in both the electron (blue) and hole regime (red). The aforementioned electron-hole asymmetry manifests here in several ways: Firstly, we observe a difference in the threshold voltage, with the transistor switching on at $V_T\simeq 1.2\;\mathrm{V}$ for $n$ and $V_T\simeq -4.2\;\mathrm{V}$ for $p$ doping. The overall transconductance, i.e. the increase of conductance with respect to gate voltage, also shows an asymmetry with the $n$ region switching on faster than its $p$ counterpart. Lastly $n$ and $p$ region differ in the number of sub-threshold Coulomb blockade oscillations. These oscillations arise due to unintentionally localised charges in the disordered nano-wire FET channel. 

\begin{figure}[htbp]
	\includegraphics[width=0.9\columnwidth]{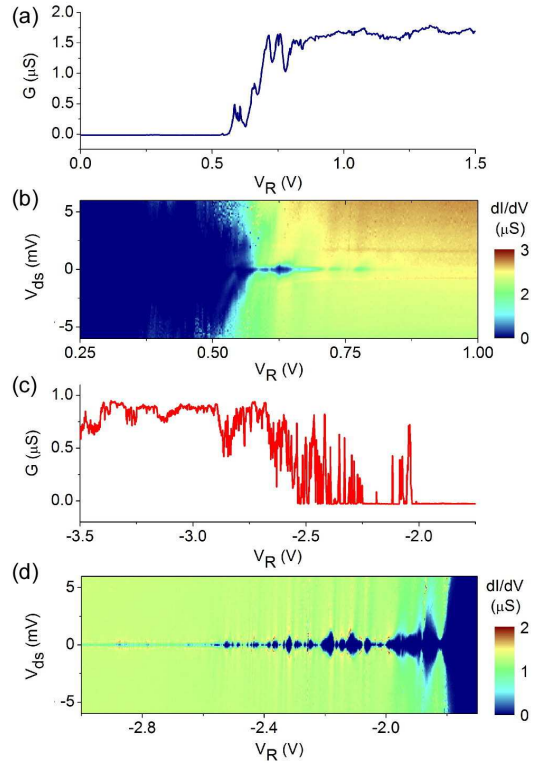}
	\caption{(Color online) (a) Conductance measurement for the right barrier in the $n$-region. The top gate voltage is $V_T=2.5\;\mathrm{V}$ (b) Differential conductance stability map of the right barrier in the $n$-region. (c) Conductance measurement for the right barrier in the $p$-region. Top gate voltage is $V_T=-4.8\;\mathrm{V}$. (d) Differential conductance map of the right barrier for $p$-type conduction. In all cases left and right barrier showed equal behaviour and only gate R is displayed.}
	\label{Fig2}
\end{figure}

We now move on to studying the effect of the tunnel barriers induced by gates L and R. The upper gate T is  set to a fixed potential, populating the channel with carriers, and we measure how the tunnel barriers beneath gates L and R switch off the channel. We restrict the description to the characteristics of tunnel barrier R as similar behaviour was noted for both tunnel barriers. Choosing $n$-type conduction, we sweep gate voltage $V_R$ for source-drain voltages $V_{ds}=\pm6\;\mathrm{mV}$. We notice several Coulomb diamonds (Fig.\ref{Fig2}(b)) or sub-threshold Coulomb oscillations (Fig.\ref{Fig2}(a)) caused by charges trapped in the barrier region. However, when the measurement is repeated in the $p$-region (Fig.s\ref{Fig2}(c) and (d)), the same tunnel barrier localises more carriers, indicated by the increased number of Coulomb diamonds in the stability map Fig.\ref{Fig2}(d) and higher number of Coulomb peaks in Fig.\ref{Fig2}(c). We attribute this increased localisation to the higher effective mass of the light hole band ($m^*\simeq 0.4m_0$) as compared to the electrons ($m^*\simeq 0.2m_0$) \cite{AndoSternRMP}. A higher effective mass means that carriers can be more easily localised by fluctuations in the 2-dimensional potential landscape.

\begin{figure}[htbp]
		\includegraphics[width=0.9\columnwidth]{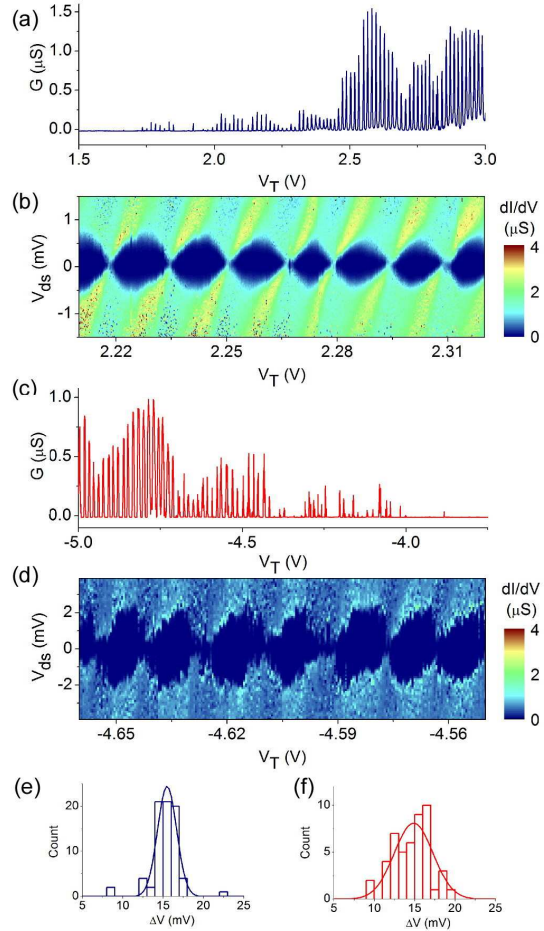}
	\caption{(Color online) Coulomb oscillations (a) and diamonds (b) measured in the $n$-region. Gates L and R were set to $V_L=0.7\;\mathrm{V}$ and $V_R=0.75\;\mathrm{V}$, respectively. Coulomb oscillations (c) and diamonds (d) in the $p$-region with gate voltages $V_L=V_R=-2.1\;\mathrm{V}$. Histogram of peak spacing for the \textit{n} (e) and \textit{p} (f) regions. }
	\label{Fig3}
\end{figure}

Finally, we measure the device configured as a quantum dot in both the $n$ and $p$ regimes. To this end, the potentials on both gates L and R are set so that tunnel barriers are formed and sequential transport of carriers takes place through the central island. As the potential on the top-gate T is swept, single period Coulomb oscillations are observed as electrons are added to the island. Fig. \ref{Fig3}(a) displays the corresponding data. We have acquired a $I_{ds}(V_{ds},T_T)$ stability map (Fig.\ref{Fig3}(b)), from which we infer a $n$-region charging energy $E_c=e^2/C_{\Sigma}\simeq 0.7{meV}$. Repeating the measurement with $p$-type conduction, we find similar behaviour for the holes (Fig.\ref{Fig3}(c)) with an approximately equal amount of oscillations. From the $p$-type stability map we find that holes are added with a charging energy of $E_c\simeq 2.4\;\mathrm{meV}$. It is difficult to make meaningful comparison of the charging energies since they vary with carrier population. We note however, that the gate period is very similar in both cases, which we have visualised in the histograms Fig.\ref{Fig3}(e) ($n$-type) and Fig.\ref{Fig3}(f) ($p$-type). Both Coulomb blockade spacing distributions can be fitted by a Gaussian revealing a mean gate spacing of $15\;\mathrm{mV}$. Furthermore, comparing Fig.\ref{Fig3}(a) and Fig.\ref{Fig3}(c), a qualitatively higher charge noise is observed for hole conduction which we attribute to redistribution of the trapped carriers within the tunnel barriers. The latter are in addition disordered, and contain more trapped carriers, in the hole regime.

In conclusion, we have fabricated and experimentally characterised ambipolar quantum dots in intrinsic silicon. We compare $n$-type and $p$-type conduction in the same device and find that while operation is successful in both regimes, the $p$-type region exhibits higher charge noise due to a larger number of localised carriers. The formation of a high-quality oxide interface in order to decrease the likelihood of carrier localisation will therefore be of particular importance when making high-quality $p$-type devices. Furthermore, the ability to fabricate and measure $p$-type quantum dots in silicon raises the possibility of measuring spin lifetime, and spin blockade in single and double quantum dots of this type. Beyond this, one can imagine an ambipolar quantum dot with a single confined electron and hole.

We thank the University of Surrey Ion Beam Centre for implanting the samples. We further acknowledge support from EPSRC Grant No. EP/H016872/1 and the Hitachi Cambridge Laboratory. A.J.F. was supported by a Hitachi Research Fellowship.


\end{document}